\documentclass[aps,preprint]{revtex4}%
\usepackage{amsfonts}
\usepackage{amsmath}
\usepackage{amssymb}
\usepackage{graphicx}%
\begin{document}
\title{Unconventional ferromagnetism and transport properties of (In,Mn)Sb dilute
magnetic semiconductor}
\author{V. N. Krivoruchko,V. Yu. Tarenkov, D. V. Varyukhin, A. I. D'yachenko}
\affiliation{Donetsk Physics \& Technology Institute NAS of Ukraine, Street R. Luxemburg
72, 83114 Donetsk, Ukraine}
\author{O. N. Pashkova, V. A. Ivanov}
\affiliation{N. S. Kurnakov Institute of General and Inorganic Chemistry of the Russian
Academy of Sciences, 31 Leninsky av., 119991 Moscow, Russia}
\date{\today}


\begin{abstract}
Narrow-gap higher mobility semiconducting alloys In$_{1-x}$Mn$_{x}$Sb were
synthesized in polycrystalline form and their magnetic and transport
properties have been investigated. Ferromagnetic response in In$_{0.98}%
$Mn$_{0.02}$Sb was detected by the observation of clear hysteresis loops up to
room temperature in direct magnetization measurements. An unconventional
(reentrant) magnetization versus temperature behavior has been found. We
explained the observed peculiarities within the frameworks of recent models
which suggest that a strong temperature dependence of the carrier density is a
crucial parameter determining carrier-mediated ferromagnetism of (III,Mn)V
semiconductors. The correlation between magnetic states and transport
properties of the sample has been discussed. The contact spectroscopy method
is used to investigate a band structure of (InMn)Sb near the Fermi level.
Measurements of the degree of charge current spin polarization have been
carried out using the point contact Andreev reflection (AR) spectroscopy. The
AR data are analyzed by introducing a quasiparticle spectrum broadening, which
is likely to be related to magnetic scattering in the contact. The AR
spectroscopy data argued that at low temperature the sample is decomposed on
metallic ferromagnetic clusters with relatively high spin polarization of
charge carriers (up to 65\% at 4.2K) within a cluster.

\end{abstract}
\maketitle

\section{Introduction}

Carrier-induced ferromagnetic order in Mn-doped III-V alloys attracts much
attention at present due to their promising combination of magnetic and
semiconducting properties (for reviews on the topic see Refs. \cite{1, 2, 3,
4}). However, the applicability of these semiconductors in microelectronic
technologies requires an increase of their Curie temperature, T$_{C}$, above
room temperature. A progress in the field at present is as large as T$_{C}$
$\approx$ 170K observed for Ga$_{1-x}$Mn$_{x}$As \cite{5}. The issue of
whether or not it is realistic to expect for (III,Mn)V materials to have
room-temperature T$_{C}$ is the key question in the field and motivates many
theoretical and experimental studies. Ferromagnetism in dilute magnetic
semiconductors (DMSs) is usually ascribed to carrier-mediated mechanisms and
depends on many different parameters (such as carrier density, magnetic
impurity density, coupling between the ion magnetic moment and the
hole/electron spin, details of disorder, etc.) which vary greatly from system
to system and even from sample to sample. According to current theoretical
models \cite{1, 2, 3, 4, 6, 7, 8, Calder07} stability of ferromagnetic state
can be enhanced by increasing carriers density in the vicinity of the magnetic
impurities. It was also demonstrated that one can control the strength of
ferromagnetic coupling (up to 25\%) by using hydrostatic pressure without any
change in the carriers concentration \cite{9}.

The picture of the ferromagnetic transition in (III,Mn)V DMSs which seems to
be accepted more or less universally is as follows. When Mn$^{3+}$ substitutes
the cation (group III element) in the lattice, the Mn absorbs an electron
converting into Mn$^{2+}$ and produces a hole. Due to strong Hund's
interaction, the half-filled 3d$^{5}$-shell Mn$^{2+}$ is spin polarized. On
the other hand, the hole experiences a Coulomb attraction to the Mn$^{2+}$
creating a spin-polarized acceptor level. This hole with ferromagnetically
ordered Mn moments form such an object as a bound magnetic polaron, which is a
localized carrier, magnetically strongly correlated with a few neighboring
magnetic moments \cite{RS}. Since the concentration of magnetic impurities is
much larger than the hole concentration, most likely due to compensation by
group V element antisite defects, a bound magnetic polaron consists of one
localized hole and a large number of magnetic impurities around the hole
localization center. Even though the direct exchange interaction between the
localized magnetic impurities is antiferromagnetic, the interaction between
bound magnetic polarons may be ferromagnetic at large enough concentrations of
holes. The ferromagnetic transition is thought to be caused by the
temperature-driven magnetic percolation transition of bound magnetic polarons
\cite{1, 2, 3, 4, 6, 7, 8, Calder07, Kam02}. Typically, the ferromagnetic
T$_{C}$ due to bound magnetic polarons percolation is relatively low. The
answer whether or not it is realistic to expect for this class of DMSs to have
Curie temperature above room temperature is still a matter of controversy
\cite{1, 2, 3, 4}.

Recently new arguments have been found which open a possibility to range the
Curie temperature in (III,Mn)V DMSs up to room temperature \cite{Calder07, 8}.
These models account for the temperature dependence of the carriers which
could, in principle, leads to a high-T carrier-mediated ferromagnetism in
semiconductors. In particular, the authors demonstrated the possibility of
stable reentrant ferromagnetism. Namely, in contrast to the standard monotonic
decay of magnetization with increasing temperature, as the temperature is
increased the higher density of thermally exited carriers can enhance the
exchange coupling between magnetic ions and thus increase the magnetization
over some temperature interval.

As already mentioned, Mn$^{2+}$ ions substituting for a trivalent cation
provide magnetic moment and act as a source of valence-band holes that mediate
the Mn$^{2+}$ - Mn$^{2+}$ interaction \cite{1, 2, 3, 4, 6, 7, 8}. This
interaction results in a ferromagnetic phase with, as expected, half-metallic
properties. The key role of spin-polarized charge carriers which mediate
ferromagnetic interaction between localized Mn$^{2+}$ ions, motivates
researchers to explore transport properties of the compounds by the point
contact (PC)\ Andreev reflection (AR) techniques. Indeed, during the last
years it was demonstrated that AR spectroscopy is an appropriate way for
direct measurement of electrical current spin polarization in a variety of
materials, including ferromagnetic metals, metallic oxides, half-metals and
semiconductors (see, e.g., Refs. \cite{30, 31, 32} and references therein).
The method is based on the difference in the AR in normal metal/superconductor
(N/S) and in ferromagnet/superconductor (F/S) contacts \cite{33}. However, it
is not an easy task to determine the electrical current spin polarization in
DMSs by the AR spectroscopy. The Schottky barrier fundamentally limits the
accuracy of spin-polarization measurements strongly decreasing the probability
of the AR. To avoid this problem, one should use heavy doped semiconductors
with metallic-type conductivity. Fortunately, for higher mobility
ferromagnetic semiconductors, such as In$_{1-x}$Mn$_{x}$Sb, the Schottky
barrier is thin, which makes the S/In$_{1-x}$Mn$_{x}$Sb interface highly
transparent and thus the AR methods applicable \cite{12, 18, 34}.

An interesting result of experimental efforts in the field, which is also
important for developing spin based devices, is discovery of a large magnitude
of charge-carriers (holes) spin polarization in filmy samples of Mn-doped
III-V semiconductors. In particular, the Ga$_{1-x}$Mn$_{x}$As is one of the
thoroughly characterized DMS and for this alloy the AR measurements have
revealed a magnitude of the carrier spin polarization up to $\approx$ 85\% for
samples with 5-8\% Mn \cite{19, 20}. Such an alloy as (In,Mn)Sb is another
relatively new ferromagnetic material, which has the largest lattice constant
in the family of (III,Mn)V materials. Since (In,Mn)Sb has the smallest
effective mass of the holes, it has much higher hole mobility than other
(III,Mn)V ferromagnetic semiconductors. It is also unique among the
ferromagnetic (III,Mn)V family due to the smallest energy gap. For In$_{1-x}%
$Mn$_{x}$Sb filmy samples charge-carriers spin polarization was found to be as
large as 52\% at liquid helium temperature \cite{18}.

In this communication, we report on the fabrication of the semiconducting
alloys In$_{1-x}$Mn$_{x}$Sb (x = 0 and 2\%Mn) in the form of large-grained
polycrystals and the results of the samples' macroscopic magnetic and
transport properties studies. Direct measurements of magnetization in
In$_{0.98}$Mn$_{0.02}$Sb were carried out and ferromagnetic order is
unambiguously established by the observation of the magnetization hysteresis
loops up to room temperature. We also observed an unconventional magnetization
\textit{vs} temperature behavior: the magnitude of the magnetization is
increased by about 3\% as temperature increase from 4.2K to room temperature.
We explain the results found based on the models of carrier-mediated
ferromagnetism in (III,Mn)V DMSs which account for the temperature dependence
of the carriers density. The correlation between magnetic states and transport
properties of the sample has been examined. Particularly, the point contact
Andreev reflection spectroscopy has been used to investigate uniformity of a
low-temperature magnetic order and charge carriers spin polarization in the
DMS under consideration. For some of the junctions the measured AR spectrum
displays characteristics which are typical for PCs of a BCS superconductor
with a high current spin polarized conductor. For these PCs the
charge-carriers spin polarization was found to be as large as 65\% at liquid
helium temperature. Apart form AR spectroscopy we also performed tunnel
spectroscopy studies and investigated a band structure of (InMn)Sb near the
Fermi level.

The paper is organized as follows. In Section II the experimental details and
some general properties of the samples are presented. Section III is devoted
to the magnetostatic data obtained. Discussion of these experimental results
and its comparison with models predictions one can find in Sec.IV. The data
for bulk transport properties of the samples are given in Sec. V. Here a
correlation between the bulk magnetic state and transport properties of the
sample has been expected, including the possibility of a simultaneous magnetic
and transport percolation \cite{22, 23, 24}. The experimental results on
Andreev reflection spectroscopy and related discussion one can find in Sec.
VI. We end with the Summary.

\section{Samples and experimental details}

A basic material, InSb, has the largest unit sell and narrowest band gap
(0.17$\div0.27$eV) among III-V semiconductors. Lightly doped InSb-Mn has also
the shallowest acceptor level among Mn-doped III-V DMSs \cite{2, 26}. However,
the equilibrium solubility of Mn in III-V compounds is quite low ($\sim$
10$^{19}$ per cm$^{3}$). Bulk samples of the solid solutions InSb-Mn (2 at.\%
Mn) were synthesized in Institute of General and Inorganic Chemistry RAS
(Moscow, Russia) as described in details earlier \cite{27}.\ In brief, the
starting materials were single crystal ISE-2 'v' n-type InSb (with electron
concentration 2$\times$10$^{14}$cm$^{-3}$ at 77K and mobility 5$\times$%
10$^{5}$cm$^{2}$/(V s)) and 99.9\% pure Mn and Sb. The polycrystalline solid
solutions In$_{1-x}$Mn$_{x}$Sb were synthesized in evacuated quartz ampoules
under isothermal conditions throughout the length of the ampoule. The samples
have been melted at 800$^{o}$C for 2h, followed by cooling to 550$^{o}$C and
solidification under nonequilibrium conditions. The crystallographic quality
of the In$_{1-x}$Mn$_{x}$Sb specimen was studied using Fe \textit{K}$\alpha$
radiation. The measured lattice constant, $a$ = 6.478 \AA , coincides with
that of InSb within the accuracy of the analysis. None of the samples studied
show any significant impurity peaks corresponding to such unwanted phases like
Mn complexes or MnSb.

To examine quantitatively the Mn$^{2+}$ distribution, the chemical composition
analysis of the In$_{0.98}$Mn$_{0.02}$Sb sample was carried out by X-ray
energy dispersive spectroscopy (EDS) using INCAPentaFETx3 spectrometer and
JSM-6490LV scanning electron microscopy (SEM). The spatial resolution of the
spectrometer in a single step is about 100nm. The crystal's fresh chip
micrograph is shown in Fig. 1. The elemental distribution of the sample was
checked using the point and shoot microanalysis at various points.
Representative data obtained from the EDS spectra for eleven points in Fig. 1
are summarized in Table I. As it follows from the EDS data, the Sb
concentration varies slightly; the concentrations of In and Mn mutually
correlate. We attribute detection of the O ions to adsorption of oxygen on the
sample's surface from air. So, we suggested that all Mn ions are in a solid
alloy state, though this alloy is somewhat inhomogeneous. Within an accuracy
of the spectrometer resolution, we did not find a decomposition of the sample
into regions with large and small concentration of magnetic ions. As it
follows from the results obtained, the average concentration of Mn$^{2+}$ ions
is 1.96(0.8)\%.

According to previous investigations \cite{27}, the In$_{1-x}$Mn$_{x}$Sb solid
solutions with x $\geq$ 1\%Mn are \textit{p}-type with a room temperature hole
concentration of $\sim$ 10$^{19}$ cm$^{-3}$. This means that manganese doping
of \textit{n}-type InSb caused a conversion to \textit{p}-type and increased
the carrier concentration by four orders of magnitude in comparison with the
pure compound semiconductor InSb (see also comments to Fig. 7 below). It is
worth to note, that \textit{the carrier concentration in the synthesized
materials is an order of magnitude greater }than the highest carrier
concentration, $\sim$ 10$^{18}$ cm$^{-3}$, in In$_{1-x}$Mn$_{x}$Sb single
crystals grown by Czochralski technique under near-equilibrium conditions and
an order of magnitude lower than the manganese concentration, $\sim$ 10$^{20}$
cm$^{-3}$, due to compensating defects.

Magnetic properties of the samples were measured by PPMS- 9 system (Quantum
Design) operated in temperature range from 4.2K to 350K under an applied
magnetic field up to 9T. The current-voltage (I \textit{vs} V) characteristics
were measured using a conventional four-probe method. Resistivity as a
function of temperature was measured directly by using an \textit{ac} voltage
bias source with a small output resistance and $\sim$ 400 $\mu$V amplitude of
signal on the sample. The tunneling measurements were carried out on tunnel
junctions formed by pressing a small Ag piece (counterelectrode) against the
sample. Tunnel barrier is formed by the native oxide surface on Ag. For the AR
spectroscopy, low-temperature superconductor Pb is used as a superconducting
electrode. Metallic contacts between In$_{0.98}$Mn$_{0.02}$Sb plate and
superconducting wire were formed by pressing slide-squash up a needle-shaped
superconductor by a micrometric screw against the mechanically polished
plate's surface. To record the AR spectra (dI/dV \textit{vs} V) of the point
contacts, we used 150$\div$200 $\mu$V modulating \textit{ac} voltage. The
resistance of the current and potential electrodes was R $\sim$ 10$^{-4}$
$\Omega\times$cm$^{2}$. The junctions resistance was much larger ($\sim$
1$\div$5 $\Omega$), so that the rescaling effects can be neglected. The
contacts' parameters were stable, offering a possibility to perform
measurements in wide temperature range.

In the report, the results obtained on ferromagnetic In$_{1-x}$Mn$_{x}$Sb with
x = 0.02 are compared where it is possible with the similar data for
ferromagnetic MnSb and nonmagnetic InSb.

\section{Magnetic properties: Experiment}

In Fig. 2, magnetization M(H) of In$_{0.98}$Mn$_{0.02}$Sb (main panel) and,
for comparison, of MnSb (bottom inset) prepared under the same conditions are
shown. Surprisingly, for In$_{0.98}$Mn$_{0.02}$Sb, clear hysteresis loops are
detected not only at T = 5K or 77K, but up to 300K (the highest upper
temperature available in the experiment). One can see in the figure that the
M(H) saturates at fields $\approx$ 4 kOe at T = 5K. The dependence of
magnetization on temperature in magnetic field H = 1 kOe is shown in upper
inset. The data revealed that the sample of In$_{0.98}$Mn$_{0.02}$Sb is
ferromagnetic with relatively small coercivity (anisotropy) up to room
temperature, and that the saturation magnetization is $\approx$ 1.45emu/g. The
main result here shown in the upper inset is that \textit{the magnitude of the
magnetization is increased by about 3\% as temperature increases from 4.2K to
room temperature.}

It should be noted here, that there are various experimental difficulties in
the unambiguous determination of the origin of magnetic signals in the DMS
containing a minute number of magnetic ions. A main question here is whether a
spatially uniform ferromagnetic order is a real ground state in this DMS or
not? When the concentration of magnetic impurities exceeds the solubility
limit, a variety of magnetic nanocrystals embedded into semiconductor may be
formed. In our case, as a first possibility, one can argue that the detected
hysteresis loops may be due to formation of MnSb clusters with T$_{C}$
$\approx$ 580K. For the In$_{1-x}$Mn$_{x}$Sb samples with x $=$ 1$\div$1.3\%
Mn prepared by the same group, such suggestion was earlier invented in Ref.
\cite{28}. Indeed, the equilibrium solubility of Mn in III-V compounds is
quite low and, if the Mn concentration is beyond the solubility limit, one
should expect a formation of nano-regions with different concentration of
magnetic ions. However, for the In$_{0.98}$Mn$_{0.02}$Sb sample, with
increasing temperature,\ we detected small but visible \textit{increase} of
the magnetization (see upper inset in Fig. 2) and \textit{increase} of the
coercivity field. The behavior of both is untypical for ferromagnets. In
addition, for MnSb, at the same conditions, T = 300K, we were not able to
detect the coercivity field: the M(T) curve is quite reversible (see bottom
inset in Fig. 2). We also tried to fit the M(H,T) data for the In$_{0.98}%
$Mn$_{0.02}$Sb sample at T = 4.2 K and 300K by the Langevin function and to
restore a typical size of possible ferromagnetic clusters. However, we found
that this is impossible if one assumes that the size of a cluster is fixed.

For some of the cases, ferromagnetic response may be due to formation of
Mn-rich nanoclusters (a spinodal decomposition) \cite{4, Moreno}. It is well
known that phase diagrams of a number of alloys exhibit a solubility gap in a
certain concentration range. This may lead to a spinodal decomposition into
regions with a low and a high concentration of magnetic constituent. It may
appear in a form of coherent nanocrystals embedded by the majority component.
Since spinodal decomposition does not usually involve a precipitation of
another crystallographic phase, it is not easily detectable experimentally.
Nevertheless, its presence was found in (Ga,Mn)As, where coherent zinc-blende
Mn-rich (Mn,Ga)As metallic nanocrystals led to an apparent Curie temperature
up to 360 K \cite{Moreno}. As was found, the magnetization behavior for
Mn-rich clusters is a conventional one (see, e.g., Fig.5b in Ref.\cite{Moreno}%
) while we observe unconventional M(T) behavior. Still, a possibility that a
small number of MnSb or Mn-reach nanoclusters are formed in (InMn)Sb matrix
cannot be ruled out completely.

It may be also argued that the samples are polycrystalline and ferromagnetism
could be due to a surface. Indeed, room temperature ferromagnetism has been
observed in a variety of inorganic nanoparticles although the materials are
intrinsically non-magnetic \cite{Rao09}. The point defects at the surface are
likely to be responsible for this phenomenon. However, the appearance of
ferromagnetism is detected only for small enough particles (%
%TCIMACRO{\TEXTsymbol{<} }%
%BeginExpansion
$<$
%EndExpansion
10 nm), while it disappears in the bigger particles \cite{Rao09}. For our
polycrystalline samples the ratio surface/volume is very small which casts
doubts on the origin of magnetism observed due to surface imperfections. Yet,
the role of grain boundaries remains open and experiments with another samples
of different Mn content could shed light on question whether the observed
reentrant ferromagnetism is due to polycrystallinity or not.

In Fig. 3, we compare our data with those of Ref. \cite{18}. In the figure
open circles are the magnetic field dependence of magnetization in In$_{1-x}%
$Mn$_{x}$Sb (x = 0.028) epitaxial film (with T$_{C}\approx$ 9K) at T = 50K
\cite{18}. One can see that this data is very well correlated with ours
(compare the results of Ref. \cite{18} at T = 50K and our data at T = 77K).
So, we can suggest that an unconventional M(T) dependence may be observed in
film samples too. The occurrence of robust ferromagnetism at room temperature
in bulk (InMn)Sb semiconductors with Mn up to 1.33\% was reported in Refs.
\cite{28, Stor06}. However, the authors were not able to determine whether the
magnetism is truly a bulk phenomenon. The observation also remained
unexplained. In our opinion, the observed M(T) behavior of In$_{0.98}%
$Mn$_{0.02}$Sb can be understood along the lines of the recently proposed
models \cite{Calder07, 8}. This point is further discussed below. It is worth
to note here that an unconventional M(T) behavior has already been observed in
some other DMSs \cite{cho98}. These observations were explained based on quite
different approaches.

\section{Magnetic properties: Discussion}

We apply the model \cite{8} to describe qualitatively the data in Figs. 2 and
3. As was already mentioned, a Mn impurity in In$_{1-x}$Mn$_{x}$Sb is
presented by two spin-degenerate levels: 'the deep level' and 'the shallow
level'. The deep level, when occupied by one electron, provides the impurity's
spin (S = 5/2). The shallow level plays the role of impurity's acceptor level.
The shallow, acceptor levels of the impurities are to donate, on average, less
then one hole per impurity. So, the Fermi level, $\varepsilon_{F}$, coincides
with the energy of a hole placed onto the shallow level of an impurity (see
Fig. 4).

Let us consider a hole associated with a particular Mn$^{2+}$ ion which is
confined in hydrogen-like orbital of radius r$_{H}$. As the acceptor
concentration increases, the hydrogen-like orbitals $\psi(r)=(\pi r_{H}%
^{3})^{-1/2}\exp(-r/r_{H})$ overlap to form an impurity band. At first, the
holes remain localized because of the influence of correlations and the donors
tend to form bound magnetic polarons \cite{9, 7, 8}. If the radius r$_{H}$ is
sufficiently large, overlap between a hydrogen hole and the cations within its
orbit leads to ferromagnetic exchange coupling between them \cite{Dust02}.
This interaction may be written in terms of the \textit{p-d} exchange
parameter J$_{pd}$ as: -$\sum_{i}$J$_{pd}$\textbf{S}$_{i}$\textbf{s(R}$_{i})$
, where \textbf{S}$_{i}$ is the impurity spin located at \textbf{R}$_{i}$,
\textbf{s(R}$_{i}$) is the carrier spin density at \textbf{R}$_{i}$.
(Following the main ideas of the model \cite{8} we still consider a simplified
case by taking into account only a long-range field from delocalized holes and
by ignoring a short-range field acting on the impurity spin from bound
electrons.) Since the concentration of holes is much smaller than that of
impurities, one localized hole is surrounded by many magnetic impurities.
Exchange interaction between the hole and magnetic impurities leads to their
mutual polarization when temperature is below exchange constant J$_{pd}$. This
structure, which consists of an almost completely spin-polarized hole and
magnetic impurities polarized by it, is conventionally called 'bound magnetic
polaron' \cite{RS, Kam02, Dust02}. As the density of polarons increases, the
polarons overlap forming polaron cluster with all impurities within this
cluster having their spins aligned in the same direction. Ferromagnetism
occurs when the polaron percolation threshold is overlapped and the infinite
clusters spanning whole sample appears. The option for boosting T$_{C}$
significantly is somehow to increase the hole's density in the vicinity of the
magnetic impurity. Indeed, within mean-field approximation, the impurity spins
act upon the carrier spins as an effective magnetic field $\propto J_{pd}%
N_{i}m(T)$; while the hole spins act upon the impurity spin with an effective
field $\propto J_{pd}N_{h}\nu(s,T)s(T)$.\ Here N$_{i}$\ is density of magnetic
impurities and N$_{h}$\ is hole's density, $\nu(s,T)$\ is temperature
dependent relative density of holes $\nu(s,T)=n(s,T)/N_{h}$\ where
$n(s,T)$\ is the holes density. As the result, the magnetization of the
magnetic impurities is described by%
\begin{equation}
m(T)=JB_{J}\left[  J\frac{J_{pd}N_{h}\nu(s,T)}{k_{B}T}s(T)\right]
\end{equation}
where $B_{J}(x)$ denotes the Brillouin function, while the magnetization of
conduction holes is%

\begin{equation}
s(T)=\frac{1}{2}\tanh\left[  \frac{J_{pd}N_{i}}{2k_{B}T}m(T)\right]
\end{equation}
By using Eqs. (1) and (2) one can obtain an expression for the Curie
temperature \cite{8}:%
\begin{equation}
T_{C}=J_{pd}[N_{i}N_{h}J(J+1)/12]^{1/2}\nu(0,T_{C})=T_{C}^{0}\nu(0,T_{C})
\end{equation}
where $T_{C}^{0}$ is the Curie temperature in the limit of completely ionized
donors \cite{coey05}. As it follows form Eq. (3), due to the factor $\nu
(s,T)$, an increase number of thermally excited carriers may be sufficient to
increase the exchange coupling between magnetic impurities, that is, a
ferromagnetic state may appear at higher temperature.

To describe the magnetization dependence on temperature, we assume the valence
band is separated from impurity acceptor levels by energy $\varepsilon_{d}$ in
the absence of magnetic order (see Fig. 4). For holes' relative density
dependence on temperature the expression:
\begin{equation}
n(s,T)=\frac{1}{2}N_{h}\kappa(s,T)[\sqrt{1+4/\kappa(s,T)}-1]\text{,}%
\end{equation}
is adopted \cite{8} where the function $\kappa(s,T)$ is $\kappa(s,T)=N_{c}%
/N_{h}\exp(-\varepsilon_{d}/k_{B}T)$, and $N_{c}=2.5(m^{\ast}/m_{0}%
)^{3/2}(T/300K)^{3/2}\times10^{19}cm^{-3}$ \cite{Ashctroft}. Materials
parameters were chosen from the data found in the literature for In$_{1-x}%
$Mn$_{x}$Sb alloys \cite{26, 27}. For numerical simulation, we suggested that
$N_{h}\approx1.1\times10^{19}cm^{-3}\ ,$\ $N_{i}\approx8.8\times10^{20}%
cm^{-3}$, $J_{pd}\approx200\div210eV\cdot\mathring{A}^{3}\ $,$\ |\varepsilon
_{d}|\approx15.5\div16.5meV$, $J=5/2$, $m^{\ast}\approx0.25\div0.43m_{0}$.
Fig. 5 illustrates (main panel) the temperature dependencies of the average
magnetization (per impurity atom) $M(T)=m(T)+\frac{N_{h}}{N_{i}}s(T)$ as well
as (inset) the $m(T)$ and $s(T)$ behavior which we obtain from Eqs. (1) and
(2). If at T $\longrightarrow0$ there is a small but finite carriers
concentration (for the data shown in Fig. 5 we chose $\nu(s,T\longrightarrow
0)=0.001$), there are three different solution for T$_{C}$: at low temperature
there is ferromagnetic order up to T$_{C1}$\ while the reentrant ferromagnetic
state is stable in the range T$_{C2}\leq$T$\leq$T$_{C3}$. If at T = 0 there
are no carries, ferromagnetic state exists only in the range T$_{C2}\leq
$T$\leq$T$_{C3}$ and is absent at both low and high temperatures,
corresponding to clustered (or spin-glass) and paramagnetic states, respectively.

It is noteworthy that for (III,Mn)V DMS spin-glass-like state may be an
unusual one. In general, there are two possibilities. As was already
mentioned, from our present results we cannot rule out completely a
probability that a small number of MnSb or Mn-rich nanoclusters are formed in
(In,Mn)Sb matrix with mutual off-orientation of clusters' magnetization.
Another origin of a spin-glass state is 'frozen' bound magnetic polarons. As
is well known (see, e.g., Ref. \cite{Yes81}) for system in a spin-glass state
in some of (low) temperature interval one can observe a reentrant
magnetization behavior. Taking into account transport properties (see next
section), we think that in temperature range 4.2K -- 30K the In$_{0.98}%
$Mn$_{0.02}$Sb sample is in a spin-glass state. In both cases (clustered or
spin-glass states) the temperature dependence of the magnetization reveals a
nonmonotonic behavior which is very different from the conventional monotonic
decay. Returning to Fig. 5, the curves look more exaggerated than the
experimental ones, but one should take into account that (i) we have about two
Mn ions on hundred unit cells, and (ii) a noticeable part of Mn ions located
outside the ferromagnetically ordered regions\ \cite{23, 24, Cs05} (see
discussion in Secs. VI and V) and are not involved into the machanism under consideration.

The mechanism described is also able to explain, at least qualitatively, our
observation of coercivity field increasing. Indeed, while the origin of
magnetic anisotropy in DMSs still remains not well understood (see recent
publications on the topic \cite{Gl, Step} and references therein), the
experimental results yield for Mn-doped DMS the linear dependence of the
uniaxial anisotropy to be proportional to the holes concentration \cite{Gl}.
So, the thermal activation of carriers may enhance as the exchange coupling
between magnetic impurities as well as the magnetic anisotropy. Unfortunately,
values of the basic parameters of the model, $J_{pd}$ and $\varepsilon_{d}$,
are not known exactly at present, and their scatter leads to both low and room
T$_{C}$ values.

The scenario above assumed the conduction band is empty. Actually, the
carriers may be thermally activated from the valence to the conduction band as
well. For mediated by thermally excited free carriers, the effective magnetic
interaction is suggested to be of the Ruderman-Kittel-Kasuya-Yosida (RKKY) -
type \cite{PR79}. Since the carrier density is low, the first node of the
oscillating RKKY function occurs at a length scale larger than the average
Mn-Mn distance, giving rise to a net ferromagnetic interaction (for details
see Ref.\cite{6}). Typically the Curie temperature due to the RKKY mechanism
is relatively high. As was pointed recently by Calder\'{o}n and Das Sarma
\cite{Calder07}, in some DMSs ferromagnetism may appear due to \textit{both}
the exchange coupling of a (localized) carrier in the \textit{impurity }band
with a few neighboring magnetic impurities (the bound magnetic polarons
percolation mechanism) and indirect exchange induced by thermally activated
carries in an otherwise empty \textit{conduction} band (the so-called
'activated' RKKY mechanism). In this case, in a sample thermal activation
leads to a high-T ferromagnetism due to the RKKY free-carrier mechanism, and
to a low-T ferromagnetism due to localized bound carriers in an impurity band
through the polaron percolation mechanism. Based on our data we cannot rule
this possibility out.

\section{Transport properties}

In Fig. 6 low temperature\ zero-field resistivity for ferromagnetic
In$_{0.98}$Mn$_{0.02}$Sb sample is shown. For comparison, in the inset, the
temperature dependence (up to room temperature) of resistivity for the
In$_{0.98}$Mn$_{0.02}$Sb and InSb prepared under the same conditions is also
displayed. As one can see in the inset of Fig. 6, addition of Mn ions results
in changing the transport properties from semiconducting to 'weakly metallic'.
Note that here the term 'metallic' we refer to the scope of the resistivity
\textit{vs} temperature and the term 'weakly' used in the sense that the
resistivity does not exhibit strong temperature dependence. At liquid nitrogen
temperature the resistivity is $\rho_{(InMn)Sb}(77K)=4.2\times10^{-4}$
$\Omega\cdot cm$, corresponding to a metallic mean free path of $k_{F}l\sim50$.

For system with reentrant ferromagnetic behavior, one can expect that
different character of the magnetic order at low-, T $\leq$ T$_{C2}$, and at
high-, T$_{C2}$ $\leq$ T $\leq$\ T$_{C3}$, temperatures will be revealed in a
different system's transport properties, too.\ Indeed, according to Monte
Carlo simulations (see, e.g. \cite{23, 24}), in sample containing a few
percent of Mn ions, small ferromagnetic regions (clusters) begin to develop
even above the mean-field Curie temperature where the local Mn density exceeds
the average. As T decreases these regions increase and may form a percolation
transport network (i.e., metallic conductivity) at T$_{C1}$, if ferromagnetic
state is realized below T$_{C1}$, or the system remains in a
clustered/spin-glass state (with activated character of conductivity) down to
T = 0. In both cases, a fraction of the spins still remain outside the
network/clusters providing charge carrier magnetic scattering \cite{23, 24,
Cs05}. Figure 6 shows that in temperature interval from 30K to 300K the
resistance increases with increasing temperature. Careful inspection of the
data in Fig. 6 (main panel) reveals that at low temperature 4.2K
%TCIMACRO{\TEXTsymbol{<} }%
%BeginExpansion
$<$
%EndExpansion
T
%TCIMACRO{\TEXTsymbol{<} }%
%BeginExpansion
$<$
%EndExpansion
30K the data are too noisy. So that, we cannot, unfortunately, indisputably
conclude whether the system has to have activated-like conductivity or
metallic one. A monotonic increase of the magnetization observed is in favor
of a spin-glass state \cite{Yes81}. In the next section, we will analyze this
temperature region more carefully by the AR spectroscopy.

Above T$_{C1}$, as temperature is \textit{increased} the metallic sate may be
realized due to thermally exited carries. In the metallic ferromagnetic phase,
T$_{C2}$
%TCIMACRO{\TEXTsymbol{<} }%
%BeginExpansion
$<$
%EndExpansion
T
%TCIMACRO{\TEXTsymbol{<} }%
%BeginExpansion
$<$
%EndExpansion
T$_{C3}$, hole mobility may be explained on the basis of a simple
band-transport picture, which facilitates the ideal limit of spin-polarized
transport. Typical temperature dependent processes giving rise to resistivity
are the electron-electron scattering, $\sim T^{2}$, and magnetic scattering on
Mn$^{2+}$ ions,\ (for spin-polarized carriers the one-magnon scattering
process is forbidden\ and two-magnon scattering gives $\sim T^{9/2}$
\cite{ko72}). In accordance with this picture, from the data in Fig. 6, one
can find that at T
%TCIMACRO{\TEXTsymbol{>} }%
%BeginExpansion
$>$
%EndExpansion
30K the resistivity indeed slowly increases as temperature increases up to the
highest of measured temperature 300K. Again, because the carrier density is
low the temperature behavior is 'weakly metallic' in the sense that the
resistivity does not exhibit strong temperature dependence.

Tunneling conductance G(V) = dI/dV for a sample -- insulator (I) -- tip tunnel
junction is a direct method for measuring of the single particle density of
state (DOS) of the sample at energy near the Fermi level $\varepsilon_{F}%
$\ \cite{Wolf}. To investigate the changes in DOS due to doping, we have
performed electron -- tunneling experiments on InSb--I--Ag and In$_{0.98}%
$Mn$_{0.02}$Sb--I--Ag tunnel junctions. Representative tunnel spectra obtained
are shown in Fig. 7. The conductance of the InSb-- I--Ag contacts demonstrates
a well-defined forbidden gap in the DOS between the valence and conductivity
bands. The width of this gap W = 0.27eV is in good accordance with the gap
which has been restored earlier from electric measurements \cite{Seeg}. For
diluted system we detected two main differences: (i) the forbidden gap is
absent and (ii) the chemical potential is now shifted corresponding to holes
type of conductivity (positive sign of voltage). These features are well
reproduced and are detected on large number of In$_{0.98}$Mn$_{0.02}$Sb--I--Ag
tunnel junctions. Taking into account our EDS analyzes, we can with confidence
conclude that the main part of Mn ions are a solid solution state and form
In$_{0.98}$Mn$_{0.02}$Sb matrix unique properties.

\section{Andreev reflection spectra}

\textit{Experiment.} -- To study transport properties of (In,Mn)Sb
semiconductor in a low temperature region (T
%TCIMACRO{\TEXTsymbol{<} }%
%BeginExpansion
$<$
%EndExpansion
T$_{C2}$), we have prepared and measured a number of different PCs. Below we
present a set of data obtained on various samples by contact formation at
different positions on the sample surface. Fig. 8 exemplarily shows the
representative characteristics of the Pb--In$_{0.98}$Mn$_{0.02}$Sb contact.
Here, the I--V dependence is presented in the main panel; the top inset
exhibits the temperature dependence of the contact's resistance R(T), and the
bottom inset illustrates the contact's AR spectra G(V) at T = 4.2K. As one can
see, at T
%TCIMACRO{\TEXTsymbol{<} }%
%BeginExpansion
$<$
%EndExpansion
T$_{SC}$(Pb) = 7.2K a sharp drop of the contact's resistivity is detected.
Reduction of the resistance for the contact under consideration is about 50\%,
a theoretical limit which one can expect for a perfect S/N junction, and an
excess current I$_{exc}$ is definitely observed. The current-voltage relation
above the superconducting transition temperature is linear, providing
additional support for the absence of the Schottky barrier. Any features
typical for a PC of a singlet pairing superconductor with spin-polarized metal
\cite{30, 31, 32} have not been detected.

However, for a few exclusive cases we had a chance to detect the PC
characteristics which we attributed to (nanoscale) ferromagnetic clusters with
spin-polarized charge carriers. In Fig. 9, the representative characteristics
of such Pb--In$_{0.98}$Mn$_{0.02}$Sb PC are exposed. As earlier, the I--V
dependence is shown in the main panel; the top inset exhibits the temperature
dependence of the contact's resistance R(T); and the bottom inset illustrates
the contact's AR spectra at T = 4.2K. As one can see, in contrast to the case
in Fig. 8, at T
%TCIMACRO{\TEXTsymbol{<} }%
%BeginExpansion
$<$
%EndExpansion
7.2K a sharp \textit{growth} of the contact's resistivity is observed. Also,
again in contrast to the case in Fig. 8, now an excess voltage V$_{exc}$ is
observed. The almost constant V$_{exc}$ value is detected for
%TCIMACRO{\TEXTsymbol{\vert}}%
%BeginExpansion
$\vert$%
%EndExpansion
V%
%TCIMACRO{\TEXTsymbol{\vert} }%
%BeginExpansion
$\vert$
%EndExpansion
$\leq$ 40mV. This proves the suggestion that heating effects could be
neglected. For comparison, we also prepared and measured (not shown) the AR
spectra for nonmagnetic InSb (the InSb -- Pb junctions). For all of the PCs of
the 'parent matrix' with Pb any peculiarities which may be attributed as due
to spin-polarized current have not been detected.

\textit{Modeling.} -- As already mentioned, the AR experiments provide a
direct measure of the current spin polarization, P = (j$_{\uparrow}$ -
j$_{\downarrow}$)/j$_{tot}$, where j$_{\uparrow(\downarrow)}$ is a partial
current of carriers with spin 'up'\ ('down') and j$_{tot}$ = (j$_{\uparrow}$ +
j$_{\downarrow}$). Indeed, the current through a ferromagnet -- superconductor
interface is determined by the charge conversion of a Cooper pair into
individual electrons. As a Cooper pair consists of two electrons with opposite
spins, the conversion is suppressed if there is spin discrimination of the
bands. So that P can be deduced from the voltage dependence of the conductance.

The spectra obtained were fitted using the models \cite{30, 31, 32}. The
diameter, \textit{d}, of a point contact with resistance R$_{N}$ was
determined for Sharvin-Wexler formula \cite{35}:%

\begin{equation}
R_{N}=\frac{4}{3}\pi d^{2}\rho\ell+\frac{\rho}{2d}\text{.}%
\end{equation}
Here $\ell$ is the mean free path of the charge carriers, and $\rho$ is the
bulk resistivity of the material. For $\rho$(T=4.2K) $\sim$ 10$^{-4}$ $\Omega$
cm and a junction's normal state resistance R$_{N}$ $\sim$ 1$\div$3 $\Omega$
the range of PC diameter values is d $\geq$ 300\AA ; we took for the mean free
path its typical value $\ell$ $\sim$ 100 \AA \ \cite{2, 3, 4}. This means that
for In$_{0.98}$Mn$_{0.02}$Sb PCs we deal with the diffusive regime of
conductivity $\ell$
%TCIMACRO{\TEXTsymbol{<}}%
%BeginExpansion
$<$%
%EndExpansion%
%TCIMACRO{\TEXTsymbol{<} }%
%BeginExpansion
$<$
%EndExpansion
\textit{d}. It is also possible that a contact consists of a number of smaller
size contacts connected in parallel. In this case each (small) contact is most
probably in the ballistic regime of conductivity ($\ell$\textit{
%TCIMACRO{\TEXTsymbol{>}}%
%BeginExpansion
$>$%
%EndExpansion%
%TCIMACRO{\TEXTsymbol{>} }%
%BeginExpansion
$>$
%EndExpansion
d}). Taking this possibility into account, the data was analyzed in terms of
the diffusive and ballistic models. We find that for high quality transparent
contacts both approaches lead to results within the scatter of the
experimental data (the authors of Ref. \cite{12} have also arrived to similar
conclusions\ analyzing AR experiments on Nd--(In,Mn)Sb PCs).

The total conductance G(V) is given by:%

\begin{equation}
G(V)=(1-P)G_{N}(V)+PG_{P}(V)\text{,}%
\end{equation}
where G$_{N}$ is the unpolarized conductance and G$_{P}$ is the fully
spin-polarized conductance. The G$_{P}$(V) is zero for energies, E
%TCIMACRO{\TEXTsymbol{<} }%
%BeginExpansion
$<$
%EndExpansion
$\Delta$, and for energies
%TCIMACRO{\TEXTsymbol{\vert}}%
%BeginExpansion
$\vert$%
%EndExpansion
E%
%TCIMACRO{\TEXTsymbol{\vert} }%
%BeginExpansion
$\vert$
%EndExpansion
%TCIMACRO{\TEXTsymbol{>} }%
%BeginExpansion
$>$
%EndExpansion
$\Delta$ is given by%

\begin{equation}
G_{P}(V)=4\beta/[(1+\beta)^{2}+4Z^{2}]\text{,}%
\end{equation}
with $\beta$ = (E + i$\Gamma$) /\{$\Delta^{2}$ -- (E + i$\Gamma$)$^{2}%
$\}$^{1/2}$. Here $\Delta$ is the superconductor energy gap, the interfacial
scattering strength is measured with a dimensionless parameter Z, and a
parameter $\Gamma$ is introduced to take into account a quasiparticle's finite
lifetime. The physical meaning of this phenomenological parameter is the
enhanced probability of inelastic scattering in the diffusive regime of
conductivity (see Ref. \cite{36} for details).

Figure 10 shows the results of fitting to the AR spectra for the conductance
of the spin-polarized region of the In$_{0.98}$Mn$_{0.02}$Sb being in contact
with Pb (see bottom inset in Fig. 9). The fitting parameters are: $\Delta$ =
1.3 meV, Z = 0.1, $\Gamma$ = 0.1 meV, and P = 0.65. The fits result in Z
values close to zero (metallic contacts) and high spin polarization of the
ferromagnetic region in In$_{0.98}$Mn$_{0.02}$Sb.

According to the results of this section, there are a few peculiarities which
definitely point to some features which should be addressed in future
investigations. As already mentioned, in the standard description, the
ferromagnetism of a cluster is mediated through holes in the valence band
\cite{22, 23, 24, Cs05} which can freely move \textit{only inside} this
cluster. In contrast, the surrounding matrix seems to lack freely moving
holes. However, our AR spectroscopy data show that a noticeable part of
Mn$^{2+}$ spins still remain outside of these ferromagnetic clusters and the
\textit{clustered state} possesses \textit{metallic} properties. Based on the
results, we conclude that the spin-flip scattering on Mn ions that are outside
the ferromagnetic clusters is the main source for magnetic depolarization of
spin-polarized holes (this possibility was recently suggested by Csontos
\textit{et al}. \cite{Cs05}). That is why the main part of the junctions under
consideration has revealed properties typical for the PCs of a conventional
superconductor with nonmagnetic metal. Note, that nano-scale regions with high
spin-polarization of charge carriers have been directly detected in (In,Mn)Sb
by Geresdi \textit{et al}. \cite{12}. The authors also found a striking
difference between the temperature dependence of the local spin polarization
and of the macroscopic magnetization, and demonstrated that nano-scale
clusters with magnetization close to the saturated value are present even well
above the magnetic phase transition temperature.

Although we have used a polycrystalline In$_{0.98}$Mn$_{0.02}$Sb sample, the
data obtained may not be the results of averaging. Due to the small contact
size the contact's characteristics is measured from a singe grain. So that,
the AR measurements have directly revealed that at low temperature sample with
2\%Mn is decomposed into metallic nano-scale regions with the carrier spin
polarization up to $\approx$ 65\%.

\section{Summary}

In this report, a narrow-gap higher mobility semiconducting alloys In$_{1-x}%
$Mn$_{x}$Sb (x = 0 and 2.0\%) were synthesized in polycrystalline form. The
magnetostatic measurements have been used to probe magnetic properties of
these DMSs. We observed an unconventional magnetization \textit{vs}
temperature behavior. Namely, the magnitude of the magnetization is increased
as temperature increases in temperature range 4.2 $\div$ 300 K. Wile we cannot
at the moment rule out completely a possibility that a small number of MnSb or
Mn-reach nanoclusters are formed in (InMn)Sb matrix and affect the
magnetization, we believe that reentrant ferromagnetism we observe is due to
(InMn)Sb matrix's properties. In our opinion, the results are consistent with
the recently proposed models \cite{8, Calder07}: for carrier-mediated
ferromagnetism of DMSs, the increasing temperature leads to an increased
number of thermally excited carriers which may be sufficient to increase the
exchange coupling between magnetic impurities and thereby to increase the
magnetization over some range of temperature. The point contact Andreev
reflection spectroscopy has been used to probe charge carriers spin
polarization in In$_{0.98}$Mn$_{0.02}$Sb at low temperature. Mostly, the
conventional Andreev reflection with an excess current and almost doubling of
the contact's normal state conductance has been detected. But in a few cases
we detected the PC characteristics which can be attributed to nanoscale
ferromagnetic ordered regions with spin-polarized charge carriers. For these
nano-clusters a magnitude of the carrier spin polarization may be as large as
up to 65\%. The results can be explained by the fact that at low temperature
the samples are decomposed on metallic ferromagnetic regions with high current
spin polarization. The findings constitute that ferromagnetism in In$_{1-x}%
$Mn$_{x}$Sb materials may happen to be strong. However, whether such a
possibility can be realized depends on many parameters and the quest remains
highly nontrivial.

The authors would like to thank V.I. Kamenev for X-ray analysis. We are also
grateful to T.E. Konstantinova and V.V. Burkhoveckii for sample's EDS analysis
and useful discussion. The work is partially supported by RFBR (Project \#
05-02-17666) and NASU under the Program 'Nanosystems, Nanomaterials, Nanotechnologies'.

\begin{center}
Figure Captions
\end{center}

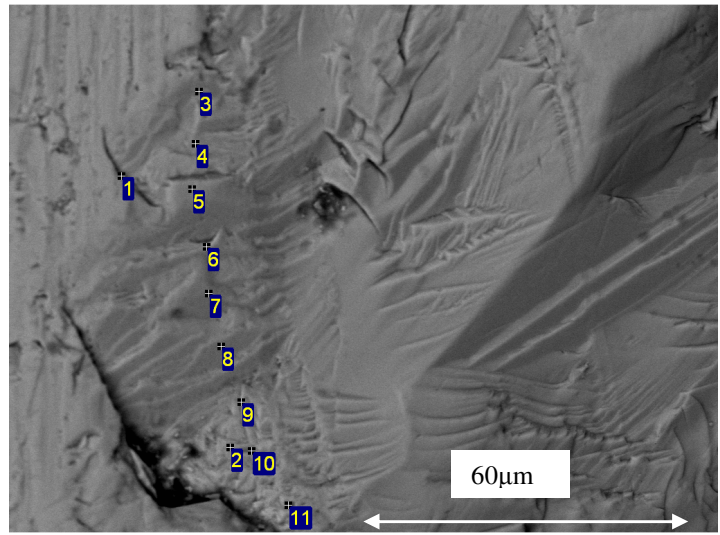
FIG. 1. SEM image of In$_{0.98}$Mn$_{0.02}$Sb sample. For numbered points the
results of the chemical composition analysis are summarized in Table I.

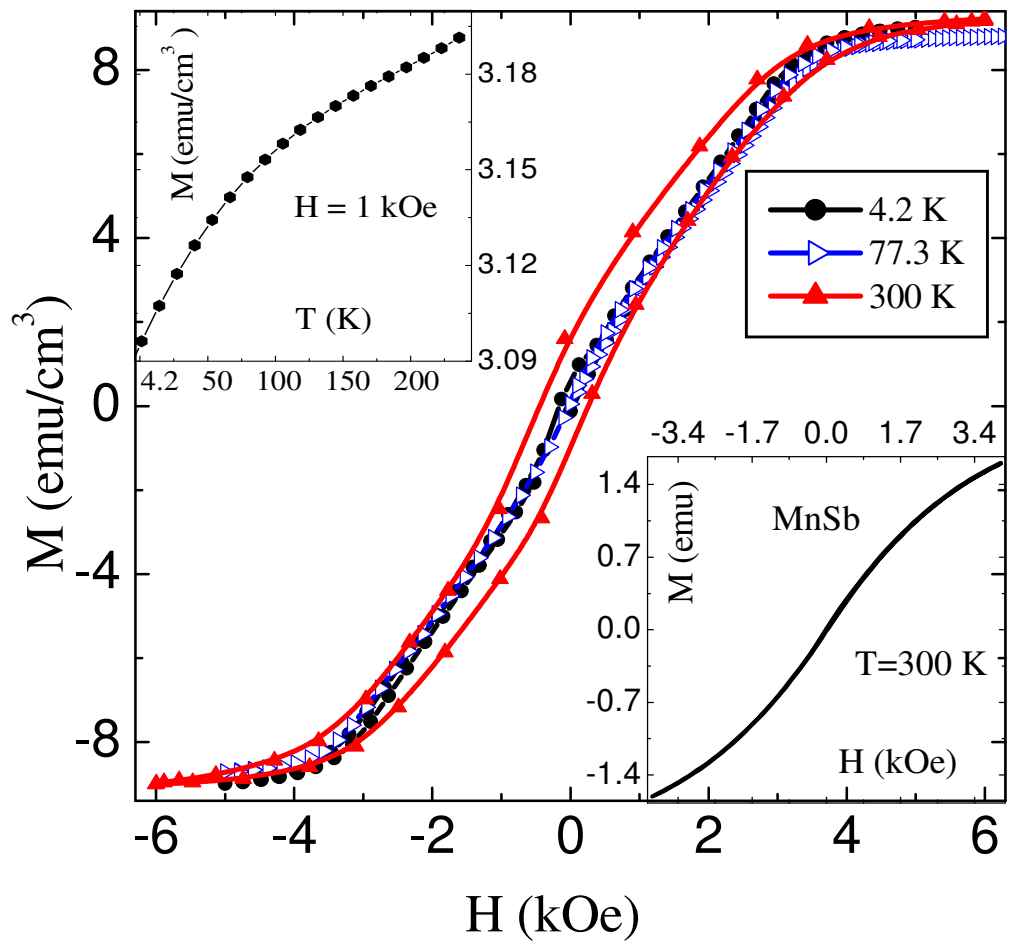
FIG. 2. (Color online) Field dependence of In$_{0.98}$Mn$_{0.02}$Sb bulk
sample magnetization; T = 4.2K (black), T = 77.3K (blue) and T = 300K (red)
curves. Upper inset: temperature dependence of In$_{0.98}$Mn$_{0.02}$Sb
magnetization at H = 1 kOe. Bottom inset: field dependence of MnSb sample
magnetization; T = 300K.

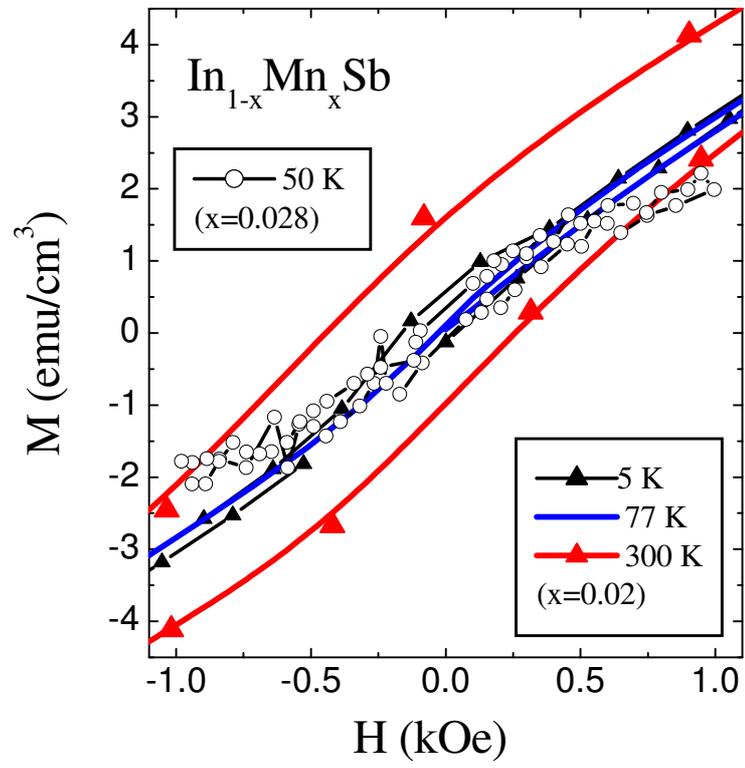
FIG. 3. (Color online) Low magnetic field magnetization dependence in
In$_{0.972}$Mn$_{0.028}$Sb (x = 0.028) epitaxial film (open circles) \cite{18}
and in In$_{0.98}$Mn$_{0.02}$Sb bulk sample.\ 

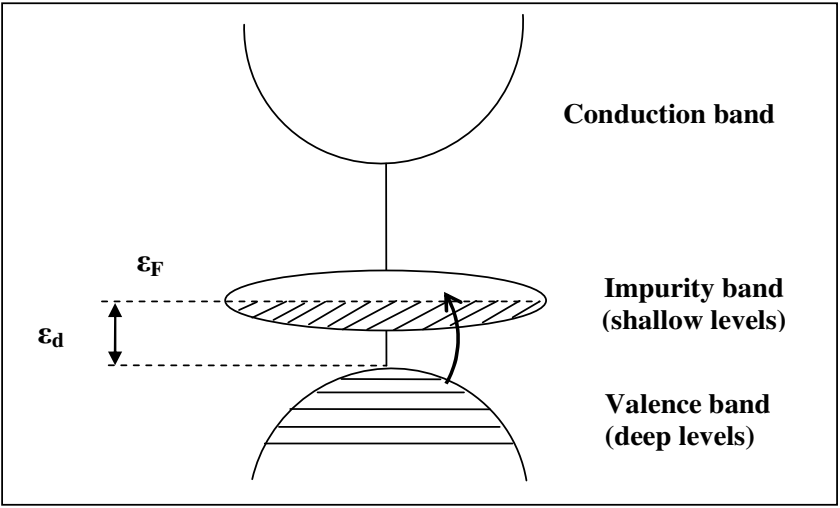
FIG. 4. Schematic diagram of the electronic structure in a hole-doped DMS.

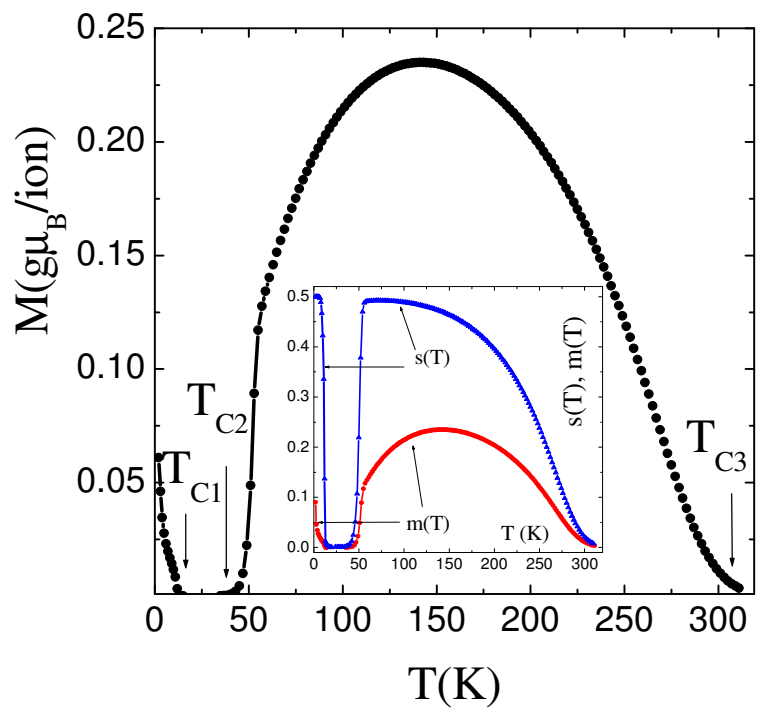
FIG. 5. (Color online)\ Temperature dependence of the average total
magnetization per ion $M(T)=m(T)+\frac{N_{h}}{N_{i}}s(T)$, showing reentrant
behavior at T = T$_{C2}$ as temperature is raised. Inset: temperature
dependence of the average spin of conduction electron s(T) and spin of
impurity ion m(T). For material's parameters see text.

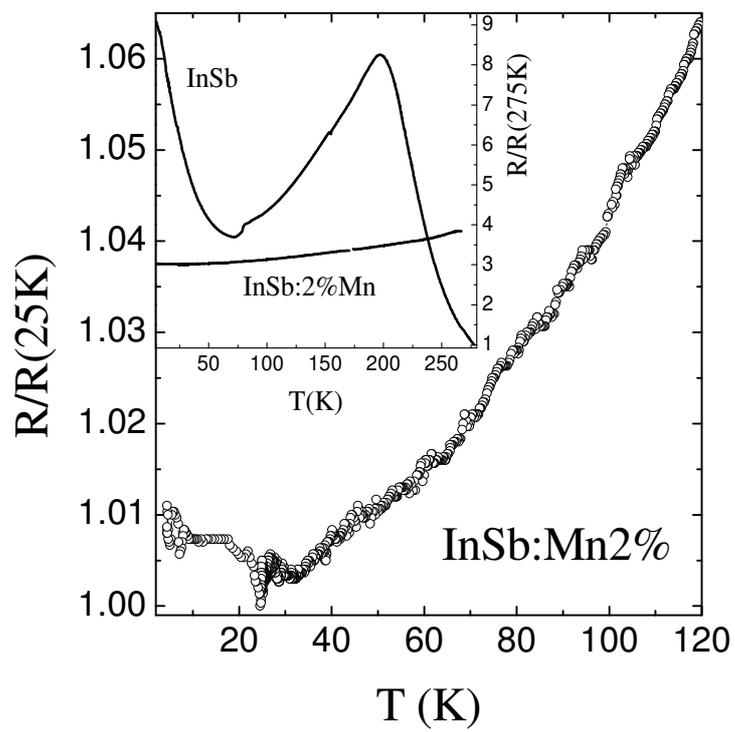
FIG. 6.\ Low-temperature dependence of the In$_{0.98}$Mn$_{0.02}$Sb
resistivity; the curve is normalized on $\rho_{(InMn)Sb}(25K)=4.09\times
10^{-4}\Omega\cdot cm$. Inset: comparative temperature dependences of the
resistivity for In$_{0.98}$Mn$_{0.02}$Sb and InSb; the curves are normalized
on $\rho_{InSb}(275K)=2.24\times10^{-2}$ $\Omega\cdot cm$.

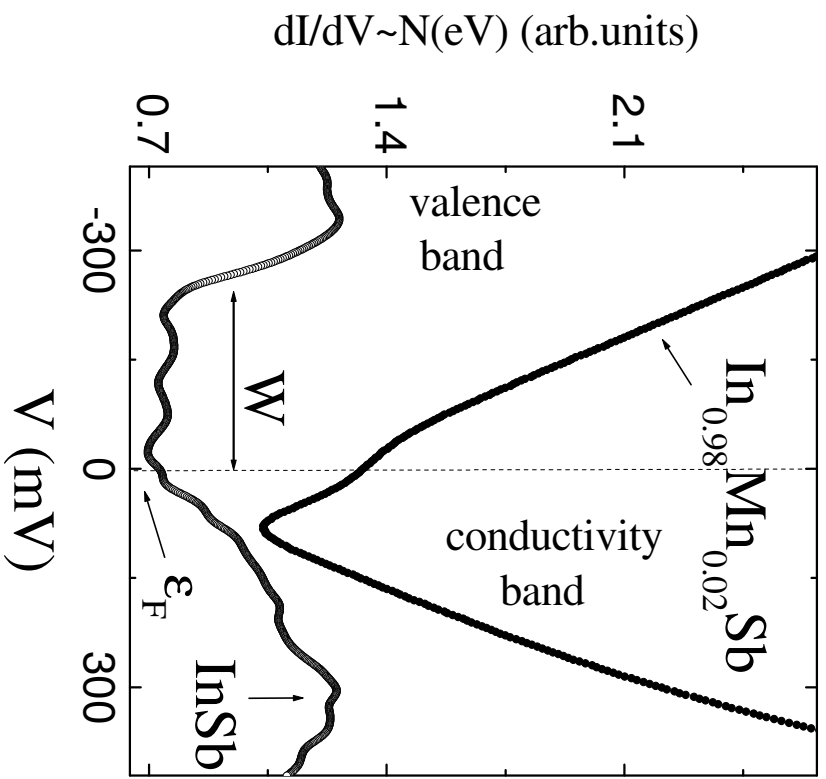
FIG. 7. Experimental density of state near the Fermi $\varepsilon_{F}$ of InSb
and In$_{0.98}$Mn$_{0.02}$Sb at T = 4.2K.

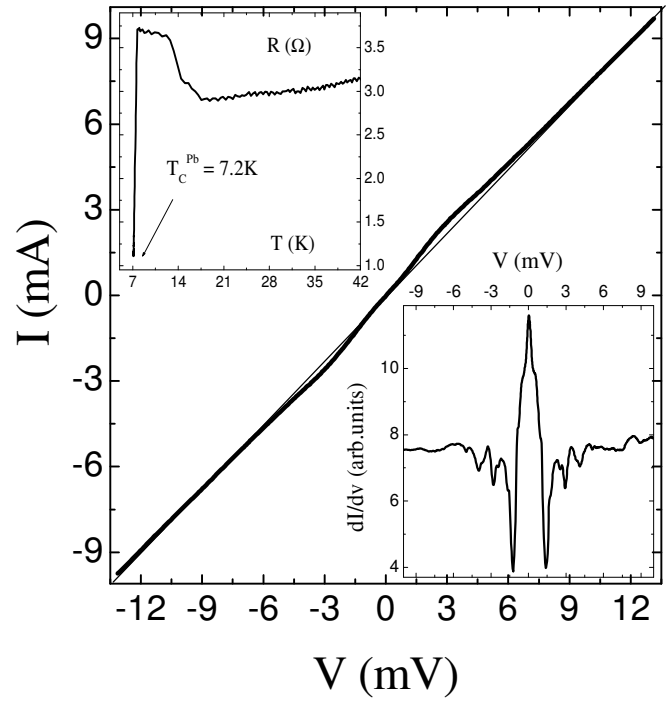
FIG. 8. Typical current-voltage (I--V) dependence of the Pb -- In$_{0.98}%
$Mn$_{0.02}$Sb contact at T = 4.2K. Upper inset: low temperature region
dependence of the contact's resistance R(T). Bottom inset: the contact's
Andreev reflection spectra at T = 4.2K.

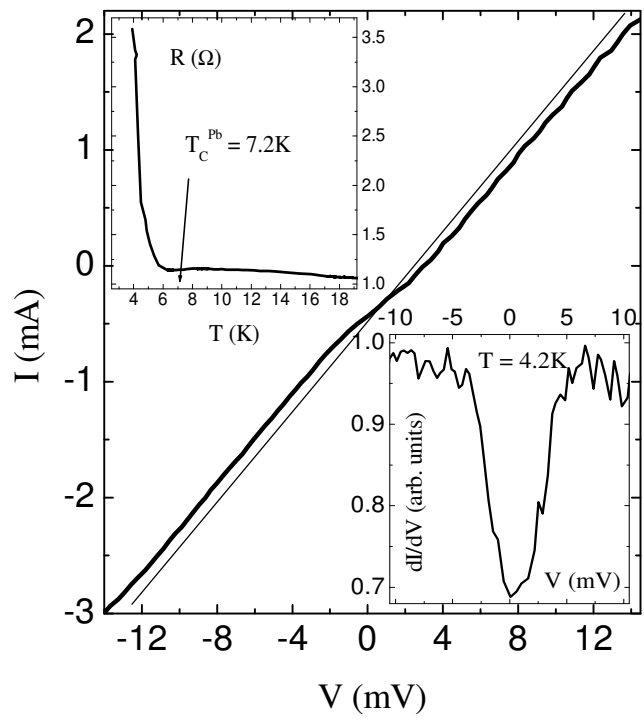
FIG. 9. An example of the current-voltage (I--V) dependence of the Pb --
In$_{0.98}$Mn$_{0.02}$Sb contact at T = 4.2K in the case when S tip is in
contact with ferromagnetic ordered nano-region of In$_{0.98}$Mn$_{0.02}$Sb.
Top inset: low temperature region dependence of the contact's resistance R(T).
Bottom inset: the contact's Andreev reflection spectra at T = 4.2K.

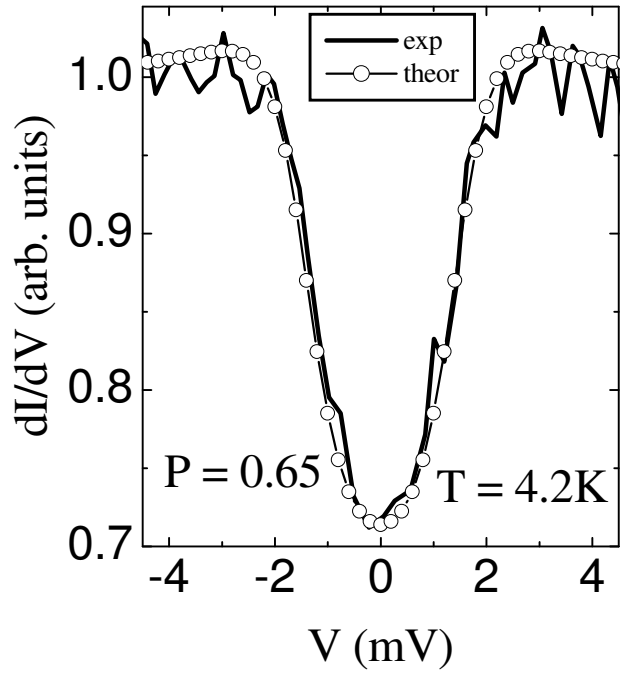
FIG. 10. Fits to modified theory (see text) for the conductance spectrum of Pb
-- In$_{0.98}$Mn$_{0.02}$Sb point contact at T = 4.2K; fitting parameters are
$\Delta$ = 1.3 meV, Z = 0.1, $\Gamma$ = 0.1 meV, and P = 0.65.\newpage

Table I. EDS data of the In$_{0.98}$Mn$_{0.02}$Sb\ sample chemical composition.%

\begin{tabular}
[c]{|c|c|c|c|c|}\hline
point%
%TCIMACRO{\TEXTsymbol{\backslash}}%
%BeginExpansion
$\backslash$%
%EndExpansion
ion & \textbf{Mn} & In & Sb & O\\\hline
1 & \textbf{1.17} & 46.65 & 50.73 & 1.46\\\hline
2 & \textbf{2.44} & 43.64 & 52.53 & 1.40\\\hline
3 & \textbf{1.50} & 45.67 & 51.22 & 1.61\\\hline
4 & \textbf{1.67} & 45.84 & 50.07 & 2.42\\\hline
5 & \textbf{3.39} & 44.11 & 51.96 & 0.54\\\hline
6 & \textbf{1.50} & 47.29 & 50.49 & 0.72\\\hline
7 & \textbf{1.69} & 45.89 & 50.99 & 1.43\\\hline
8 & \textbf{1.83} & 45.74 & 51.54 & 0.89\\\hline
9 & \textbf{2.78} & 42.69 & 51.33 & 3.20\\\hline
10 & \textbf{1.33} & 45.57 & 52.08 & 1.02\\\hline
11 & \textbf{1.96} & 45.60 & 50.81 & 1.62\\\hline
average & \textbf{1.96} & 43.91 & 51.68 & 2.45\\\hline
standard deviation & \textbf{0.80} & 5.78 & 1.99 & 3.44\\\hline
\end{tabular}

\end{document}